\documentclass[journal,letterpaper,12pt,oneside,onecolumn,draftclsnofoot]{IEEEtran}
\usepackage{amsfonts}

\usepackage[dvips]{graphicx}
\usepackage{amsmath}
\usepackage{amssymb}
\usepackage{multirow}
\usepackage{longtable}
\usepackage{subfigure}
\usepackage{float}
\usepackage{afterpage}
\usepackage{multirow}
\usepackage[nospace]{cite}
\usepackage{subeqnarray}
\usepackage{algorithm}
\usepackage{algpseudocode}
\usepackage{mathrsfs}
\usepackage{xcolor}
\long\def\symbolfootnote[#1]#2{\begingroup
\def\thefootnote{\fnsymbol{footnote}}
\footnote[#1]{#2}\endgroup}
\IEEEoverridecommandlockouts

\begin{document}
\title{\Large \bf State Transition Analysis of Time-Frequency Resource Conversion-based Call Admission Control for LTE-Type Cellular Network
%\thanks{This work was supported by National Natural Science Foundation of China (No. 61201228), Specialized Research Fund for the Doctoral Program of Higher Education (No. 20120101120077), and Zhejiang Provincial Natural Science Foundation of China (No. LY12F01021).}
}
%\vspace{-0.8cm}
\author{\small \authorblockN{Hangguan Shan$^{\dag}$, Zhifeng Ni$^{\dag}$, Weihua Zhuang$^{\S}$, Aiping Huang$^{\dag}$, and Wei Wang$^{\dag}$}\\%
\authorblockA{$^{\dag}$Department of Information Science and Electronic Engineering,
      Zhejiang University, Hangzhou, China
      }
\authorblockA{$^{\S}$Department of Electrical and Computer Engineering,
        University of Waterloo, Canada
      }
      %\vspace{0.05cm}
}

\maketitle
%\symbolfootnote[0]{$^\dagger$This work was supported by National Natural Science Foundation of China (No. 61201228), Specialized Research Fund for the Doctoral Program of Higher Education (No. 20120101120077), and Zhejiang Provincial Natural Science Foundation of China (No. LY12F01021).}

\vspace{-0.30cm}
\begin{abstract}

To address network congestion stemmed from traffic generated by advanced user equipments, in \cite{Shan_TWC_submitted} we propose a novel network resource allocation strategy, time-frequency resource conversion (TFRC), via exploiting user behavior, a specific kind of context information.  Considering an LTE-type cellular network, a call admission control policy called double-threshold guard channel policy is proposed there to facilitate the implementation of TFRC.
In this report, we present state transition analysis of this TFRC-based call admission control policy for an LTE-type cellular network. Overall, there are five categories of events that can trigger a transition of the system state: 1)  a new call arrival; 2) a handoff user arrival; 3) a handoff user departure; 4) a call termination; and 5) a periodic time-frequency resource conversion. We analyze them case by case in this report and % and present an approach to derive handoff rates of different kinds of users in the network.
the validation of the analysis has been provided in \cite{Shan_TWC_submitted}.

\end{abstract}

%\vspace{0.3cm}
%%%%%%%%%%%%%%%%%%%%%%%%%%%%%%%%%%%%
% Introduction
%%%%%%%%%%%%%%%%%%%%%%%%%%%%%%%%%%%%
%\section{Introduction}
%\label{sec:introduction}
%
%Mobile data traffic has been grown in a explosive way. Explosive traffic
%

\section{State Transitions}
\label{sec:Analysis:transition}

The events that trigger a transition of the system state defined in \cite{Shan_TWC_submitted} can be classified into five categories: 1) a new call arrival; 2) a handoff user arrival; 3) a handoff user departure; 4) a call termination; and 5) a periodic time-frequency resource conversion. In the following, we describe each category case by case. The important symbols defined in \cite{Shan_TWC_submitted} and re-utilized in this report are summarized in Table I.

\begin{table*}
\renewcommand{\arraystretch}{0.5}
\centering
\caption{Summary of important symbol.}
\vspace{-0.3cm}
\addtolength{\tabcolsep}{-4.2pt}
\begin{tabular}{|l|l||l|l|}
\hline
Symbol & Definition & Symbol & Definition \\ \hline\hline
$T_{1}$ ($T_{2}$) & Wide-band (narrow-band) connection & $U_{T_{1},T_{1}}$ &
\begin{tabular}{l}
User set in which users with two simultaneous \\
$T_{1}$ connections, one in state $S_{1}$ and the other \\
in state $S_{2}$%
\end{tabular}
\\ \hline
$\lambda _{u}$ & Connection arrival rate per user & $U_{T_{1},T_{2}}$ &
\begin{tabular}{l}
User set in which users with two simultaneous \\
connections, one $T_{1}$ connection in state $S_{2}$ and \\
one $T_{2}$ connection in state $S_{1}$%
\end{tabular}
\\ \hline
$r_{1}$ ($r_{2}$) &
\begin{tabular}{l}
Number of subchannels used by a $T_{1}$ ($T_{2}$) \\
connection%
\end{tabular}
& $U_{T_{2},T_{1}}$ &
\begin{tabular}{l}
User set in which users with two simultaneous \\
connections, one $T_{2}$ connection in state $S_{2}$ and \\
one $T_{1}$ connection in state $S_{1}$%
\end{tabular}
\\ \hline
$P_{1}$ ($P_{2}$) &
\begin{tabular}{l}
Probability of any user initiating a $T_{1}$ ($T_{2}$) \\
connection%
\end{tabular}
& $U_{T_{2},T_{2}}$ &
\begin{tabular}{l}
User set in which users with two simultaneous \\
$T_{2}$ connections, one in state $S_{1}$ and the other \\
in state $S_{2}$%
\end{tabular}
\\ \hline
$1/\mu _{1}$ ($1/\mu _{2}$) &
\begin{tabular}{l}
Mean of exponential distributed $T_{1}$'s ($T_{2}$'s) \\
connection duration%
\end{tabular}
& $m_{I}$ &
\begin{tabular}{l}
The maximum number of report times before \\
the subchannel number of a user's backgorund \\
$T_{1}$ connection decreases to 0 while the fore- \\
ground connection is another $T_{1}$ connection%
\end{tabular}
\\ \hline
$1/\eta $ &
\begin{tabular}{l}
Mean of exponentially distributed cell residual \\
time%
\end{tabular}
& $m_{II}$ &
\begin{tabular}{l}
The maximum number of report times before \\
the subchannel number of a user's backgorund \\
$T_{1}$ connection decreases to 0 while the fore- \\
ground connection is a $T_{2}$ connection%
\end{tabular}
\\ \hline
$C$ & Subchannel number in a cell & $m_{III}$ &
\begin{tabular}{l}
The maximum number of report times before \\
the subchannel number of a user's backgorund \\
$T_{2}$ connection decreases to 0 while the fore- \\
ground connection is a $T_{1}$ connection%
\end{tabular}
\\ \hline
$C_{R}$ ($C_{HR}$) &
\begin{tabular}{l}
Number of subchannels reserved for recover- \\
ing (both recoverying and handoff) calls%
\end{tabular}
& $m_{IV}$ &
\begin{tabular}{l}
The maximum number of report times before \\
the subchannel number of a user's backgorund \\
$T_{2}$ connection decreases to 0 while the fore- \\
ground connection is another $T_{2}$ connection%
\end{tabular}
\\ \hline
${\bar{K}_{A}}$ & Average user number per cell &
\begin{tabular}{l}
$U_{T_{1}^{(i)},T_{1}}$ \\
$0 \leq i \leq m_{I}$%
\end{tabular}
&
\begin{tabular}{l}
Set of users who have two simultaneous con-\\
nections in which a $T_{1}$ connection is in TFRC \\
 thus being withdrawn $r_{T_{1},T_{1}}^{(i)}$ subchannels%
\end{tabular}
\\ \hline
$R_{b}$ & Average data rate per subchannel &
\begin{tabular}{l}
$U_{T_{1}^{(i)},T_{2}}$ \\
$0\leq i\leq m_{II}$%
\end{tabular}
&
\begin{tabular}{l}
Set of users who have two simultaneous con-\\
nections in which a $T_{1}$ connection is in TFRC\\
thus being withdrawn $r_{T_{1},T_{2}}^{(i)}$ subchannels%
\end{tabular}
\\ \hline
$\tau $ & Feedback period of context information &
\begin{tabular}{l}
$U_{T_{2}^{(i)},T_{1}}$ \\
$0\leq i\leq m_{III}$%
\end{tabular}
&
\begin{tabular}{l}
Set of users who have two simultaneous con-\\
nections in which a $T_{2}$ connection is in TFRC\\
thus being withdrawn $r_{T_{2},T_{1}}^{(i)}$ subchannels%
\end{tabular}
\\ \hline
$S_{1}$ ($S_{2}$) &
\begin{tabular}{l}
Connection state in which the connection is \\
in data tansmission and in the forground \\
(background) of the screen%
\end{tabular}
&
\begin{tabular}{l}
$U_{T_{2}^{(i)},T_{2}}$ \\
$0\leq i\leq m_{IV}$%
\end{tabular}
&
\begin{tabular}{l}
Set of users who have two simultaneous con-\\
nections in which a $T_{2}$ connection is in TFRC\\
 thus being withdrawn $r_{T_{2},T_{2}}^{(i)}$ subchannels%
\end{tabular}
\\ \hline
$S_{T}$ & System state & $n_{1}(n_{2})$ & Number of users in $U_{T_{1}}$ ($%
U_{T_{2}}$) \\ \hline
$R(S_{T})$ &
\begin{tabular}{l}
Cell load measured by used subchannel number \\
given system state $S_{T}$%
\end{tabular}
&
\begin{tabular}{l}
${\cal N}_{I}^{(i)}$ \\
$0\leq i\leq m_{I}$%
\end{tabular}
& Number of users in $U_{T_{1}^{(i)},T_{1}}$ \\ \hline
$R_{rv}(j_{1},j_{2},m)$ &
\begin{tabular}{l}
Data amount connection $j_{1}$ received between   \\
the time that user $j$ pays attention to new con-  \\
nection $j_{2}$ and the time that the resource man-  \\
ager detects that the user focuses on  $j_{1}$ again%
\end{tabular}
&
\begin{tabular}{l}
${\cal N}_{II}^{(i)}$ \\
$0\leq i\leq m_{II}$%
\end{tabular}
& Number of users in $U_{T_{1}^{(i)},T_{2}}$ \\ \hline
$U_{idle}(S_{T})$ & Number of idle users given state $S_{T}$ &
\begin{tabular}{l}
${\cal N}_{III}^{(i)}$ \\
$0\leq i\leq m_{III}$%
\end{tabular}
& Number of users in $U_{T_{2}^{(i)},T_{1}}$ \\ \hline
$U_{T_{1}}$ ($U_{T_{2}}$) &
\begin{tabular}{l}
User set in which users with only one $T_{1}$ ($T_{2}$) \\
connection in state $S_{1}$%
\end{tabular}
&
\begin{tabular}{l}
${\cal N}_{IV}^{(i)}$ \\
$0\leq i\leq m_{IV}$%
\end{tabular}
& Number of users in $U_{T_{2}^{(i)},T_{2}}$ \\ \hline
\end{tabular}
\end{table*}

\subsection{A New Call Arrival}
\subsubsection{A new $T_1$-type call arrival}
When a new $T_1$-type call arrives in the cell, it will be accepted if $R({S_T}) + {r_1} \le C - {C_{HR}}$; Otherwise, the call is blocked.
Depending on the user who initiates the new call, there are three types of state transitions:
\begin{itemize}
  \item When the call is initiated by a user without any connection, ${n_1} \to {n_1} + 1$, with a rate of ${U_{idle}}\left( {{S_T}} \right){\lambda _u}{P_1}$;
  \item When the call is initiated by a user with only  a single $T_1$-type connection, ${n_1} \to {n_1} - 1$, ${\cal N}_{I}^{\left( 0 \right)} \to {\cal N}_{I}^{\left( 0 \right)} + 1$, with a rate of ${n_1}{\lambda _u}{P_1}$;
  \item When the call is initiated by a user with only  a single $T_2$-type connection,  ${n_2} \to {n_2} - 1$, ${\cal N}_{III}^{\left( 0 \right)} \to {\cal N}_{III}^{\left( 0 \right)} + 1$, with a rate of ${n_2}{\lambda _u}{P_1}$.
\end{itemize}

\subsubsection{A new $T_2$-type call arrival}
Similarly, when a new  $T_2$-type call arrives in the cell, it will be accepted if $R\left( {{S_T}} \right) + {r_2} \le C - {C_{HR}}$; Otherwise, the call is blocked. Again, depending on the user who initiates the new call, there are three types of state transitions:
\begin{itemize}
  \item When the call is initiated by a user without any connection, ${n_2} \to {n_2} + 1$, with a rate of ${U_{idle}}\left( {{S_T}} \right){\lambda _u}{P_2}$;
  \item When the call is initiated by a user with only a single $T_1$-type connection, ${n_1} \to {n_1} - 1$, ${\cal N}_{II}^{\left( 0 \right)} \to {\cal N}_{II}^{\left( 0 \right)} + 1$, with a rate of ${n_1}{\lambda _u}{P_2}$;
  \item When the call is initiated by a user with only a single $T_2$-type connection, ${n_2} \to {n_2} - 1$, ${\cal N}_{IV}^{\left( 0 \right)} \to {\cal N}_{IV}^{\left( 0 \right)} + 1$, with a rate of ${n_2}{\lambda _u}{P_2}$.
\end{itemize}
%
%
% 1) Initiated by users without any connections: ${n_2} \to {n_2} + 1$, with a rate of ${U_{idle}}\left( {{S_T}} \right){\lambda _u}{P_2}$;
% 2) Initiated by users only with $T_1$  connection: ${n_1} \to {n_1} - 1$, ${{\cal N}_4}^{\left( 0 \right)} \to {{\cal N}_4}^{\left( 0 \right)} + 1$, with a rate of ${n_1}{\lambda _u}{P_2}$; 3) Initiated by users only with $T_2$ connection: ${n_2} \to {n_2} - 1$, ${{\cal N}_6}^{\left( 0 \right)} \to {{\cal N}_6}^{\left( 0 \right)} + 1$, with a rate of ${n_2}{\lambda _u}{P_2}$.

\subsection{A Handoff User Arrival}
\subsubsection{A handoff user arrival for acceptance in $U_{T_1}$}
When a handoff user with a single $T_1$ connection arrives, ${n_1} \to {n_1} + 1$ if $R\left( {{S_T}} \right) + {r_1} \le C - {C_R}$; Otherwise, the handoff requester fails to be accepted in the cell.
According to Appendix A, handoff rate for this type of user is given by ${\lambda _{h_1}} = {\bar K_A} \cdot \eta \cdot \pi ({S_u}={r_1})$, where $\pi (S_u = x)$ is the steady-state probability of a user in the cell exactly occupying $x$ subchannels.

\subsubsection{A handoff user arrival for acceptance in $U_{T_2}$}
When a handoff user with a single $T_2$ connection arrives, ${n_2} \to {n_2} + 1$ if $R\left( {{S_T}} \right) + {r_2} \le C - {C_R}$; Otherwise, the handoff requester fails to be accepted in the cell.
Again, according to Appendix A, handoff rate for this type of user is given by ${\lambda _{h_2}} = {\bar K_A} \cdot \eta \cdot \pi ({S_u}={r_2})$. %, where $\pi (S_u = r_2)$ is the steady-state probability of a user in the cell with a connection exactly occupying $r_2$ subchannels.

\subsubsection{A handoff user arrival for acceptance in $U_{T_1, T_1}$}

A handoff user may have multiple simultaneous connections. %In case that the handoff user has two $T_1$ connections: one in state $S_1$ and the other in state $S_2$ thus been take
Such handoff user is individually considered according to the type of connections in service and the progress of the TFRC applied to the user.
%
%Handoff call associated with users in $U_{T_1^{(i)}, T_1}$ arrives with a rate (
Let $\lambda _{h( I )}^{(i)}$ denote the arrival rate of handoff users who have two $T_1$ connections in which one is in state $S_1$ and the other is in state $S_2$ and has been taken away $r^{(i)}_I$ subchannels (i.e., belonging to $U_{T_1^{(i)}, T_1}$), where $i \in \{0, 1, ...,m_I\}$.
According to Appendix A, $\lambda _{h( I )}^{(i)}$ is equal to ${\bar K_A} \cdot \eta \cdot \pi ({S_u}={2 r_1 - r^{(i)}_I})$.
When this type of user moves inward:
\begin{itemize}
  \item ${\cal N}_I^{( i )} \to {\cal N}_I^{( i )} + 1$ if $R( {{S_T}} ) + 2{r_1} - r_I^{( i )} \le C - {C_R}$ (i.e., there exist sufficient resources to admit the user's both connections);
  \item ${\cal N}_I^{( {{m_I}} )} \to {\cal N}_I^{( {{m_I}} )} + 1$ if $R( {{S_T}} ) + {r_1} \le C - {C_R} < R( {{S_T}} ) + 2{r_1} - r_I^{( i )}$ and $0 \leq i \leq m_I -1$ (i.e., the available resource can only afford one of the user's connections, thus admitting the connection with user's focus and freezing the other);
  \item Otherwise, the two connections are both dropped if $R( {{S_T}} ) + {r_1} > C - {C_R}$.
\end{itemize}

\subsubsection{A handoff user arrival for acceptance in $U_{T_1, T_2}$}
Similarly, handoff users belonging to $U_{T_1^{(i)}, T_2}$ arrive with a rate of $\lambda _{h( II )}^{(i)} = {\bar K_A} \cdot \eta \cdot \pi ({S_u}={r_1 - r^{(i)}_{II} + r_2})$, where $i \in \{0, 1, ...,m_{II}\}$.
When this type of user moves inward:
\begin{itemize}
  \item ${\cal N}_{II}^{(i)} \to {\cal N}_{II}^{(i)} + 1$ if $R(S_T) + {r_1} - r_{II}^{(i)} + {r_2} \le C - {C_R}$ (i.e., there exist sufficient resources to admit the user's both connections);
  \item ${\cal N}_{II}^{(m_{II})} \to {\cal N}_{II}^{(m_{II})} + 1$ if $R(S_T) + {r_2} \le C - {C_R} < R(S_T) + {r_1} - r_{II}^{(i)} + {r_2}$ and $0 \leq i \leq m_{II} -1$ (i.e., the available resource can only afford one of the user's connections, thus admitting the connection with user's focus and freezing the other);
  \item Otherwise, the two connections are both dropped if $R(S_T) + {r_2} > C - {C_R}$.
\end{itemize}

\subsubsection{A handoff user arrival for acceptance in $U_{T_2, T_1}$}
Handoff users belonging to $U_{T_2^{(i)}, T_1}$ arrive with a rate of $\lambda _{h(III)}^{(i)} = {\bar K_A} \cdot \eta \cdot \pi ({S_u}={r_2 - r^{(i)}_{III}} + r_1)$, where $i \in \{0, 1, ...,m_{III}\}$.
When this type of user moves inward:
\begin{itemize}
  \item ${\cal N}_{III}^{( i )} \to {\cal N}_{III}^{(i)} + 1$ if $R(S_T) + {r_2} - r_{III}^{(i)} + {r_1}  \le C - {C_R}$ (i.e., there exist sufficient resources to admit the user's both connections);
  \item ${\cal N}_{III}^{(m_{III})} \to {\cal N}_{III}^{(m_{III})} + 1$ if $R(S_T) + {r_1} \le C - {C_R} < R(S_T) + {r_2} - r_{III}^{(i)} + {r_1}$ and $0 \leq i \leq m_{III} -1$ (i.e., the available resource can only afford one of the user's connections, thus admitting the connection with user's focus and freezing the other);
  \item Otherwise, the two connections are both dropped if $R(S_T) + {r_1} > C - {C_R}$.
\end{itemize}

\subsubsection{A handoff user arrival for acceptance in $U_{T_2, T_2}$}
Handoff users belonging to $U_{T_2^{(i)}, T_2}$  arrives with a rate of $\lambda _{h(IV)}^{(i)} = {\bar K_A} \cdot \eta \cdot \pi ({S_u}={2r_2 - r^{(i)}_{IV}})$, where $i \in \{0, 1, ...,m_{IV}\}$.
When this type of user moves inward:
\begin{itemize}
  \item ${\cal N}_{IV}^{(i)} \to {\cal N}_{IV}^{(i)} + 1$ if $R(S_T) + 2{r_2} - r_{IV}^{(i)} \le C - {C_R}$ (i.e., there exist sufficient resources to admit the user's both connections);
  \item ${\cal N}_{IV}^{(m_{IV})} \to {\cal N}_{IV}^{(m_{IV})} + 1$ if $R(S_T) + {r_2} \le C - {C_R} < R(S_T) + 2{r_2} - r_{IV}^{(i)}$  and $0 \leq i \leq m_{IV} -1$ (i.e., the available resource can only afford one of the user's connections, thus admitting the connection with user's focus and freezing the other);
  \item Otherwise, the two connections are both dropped if $R(S_T) + {r_2} > C - {C_R}$.
\end{itemize}

\subsection{A Handoff User Departure}

When a user with connection(s) moves outward, a handoff user departs.
\begin{itemize}
  \item If the user is in $U_{T_1}$, ${n_1} \to {n_1} - 1$, with a rate of ${n_1}\eta$;
  \item If the user is in $U_{T_2}$, ${n_2} \to {n_2} - 1$, with a rate of ${n_2}\eta$;
  \item If the user is in $U_{T_1^{(i)},T_1}$ where $i \in \{0, 1,..., m_I\}$,  ${\cal N}_I^{(i)} \to {\cal N}_I^{(i)} - 1$, with a rate of ${\cal N}_I^{(i )}\eta$;
  \item If the user is in $U_{T_1^{(i)},T_2}$ where $i \in \{0, 1,..., m_{II}\}$,  ${\cal N}_{II}^{(i)} \to {\cal N}_{II}^{(i)} - 1$, with a rate of ${\cal N}_{II}^{(i)}\eta$;
  \item If the user is in $U_{T_2^{(i)},T_1}$ where $i \in \{0, 1,..., m_{III}\}$, ${\cal N}_{III}^{(i)} \to {\cal N}_{III}^{(i)} - 1$, with a rate of ${\cal N}_{III}^{(i)}\eta$;
  \item If the user is in $U_{T_2^{(i)},T_2}$ where $i \in \{0, 1,..., m_{IV}\}$, ${\cal N}_{IV}^{(i)} \to {\cal N}_{IV}^{(i)} - 1$, with a rate of ${\cal N}_{IV}^{(i)}\eta$.
\end{itemize}

\subsection{A Call Termination}

\subsubsection{Call terminated by users in $U_{T_1}$}
When a call belonging to a user  in $U_{T_1}$ terminates, ${n_1} \to {n_1} - 1$.
As proved in Appendix B, for a connection that has been applied time-frequency resource conversion, as long as it is successfully recovered, its call holding time after spectrum recovery follows the same distribution as the one that has not been applied TFRC. Therefore, call termination associated with users in $U_{T_1}$ occurs with a rate of  ${n_1}{\mu _1}$.
%
% (it can be proved that when the converted connection is successfully recovered, its call holding time is exponentially distributed with the same mean as before, the details are shown in Appendix A).

\subsubsection{Call terminated by users in $U_{T_2}$}
Similarly, when a call belonging to a user in $U_{T_2}$ terminates, ${n_2} \to {n_2} - 1$. And call termination associated with this type of user occurs with a rate of  ${n_2}{\mu _2}$.
%When a call belonging to a user of $U\left\{ {{T_2}} \right\}$  terminates at a rate of ${n_2}{\mu _2}$, ${n_2} \to {n_2} - 1$.

\subsubsection{Call terminated by users in $U_{T_1,T_1}$}
Specifically, when a call belonging to a user in $U_{T_1^{(i)},T_1}$ ($i \in \{0, 1,..., m_I\}$) terminates, $N_I^{(i)} \to N_I^{(i)} - 1$, ${n_1} \to {n_1} + 1$.
Since this type of user has two connections simultaneously, the call termination may be caused by either of the two connections.
As shown in Appendix A of \cite{Shan_TWC_submitted}, if the terminated connection is the one in TFRC, its service rate  is equal to ${\mu _1}( 1-r^{(i)}_I / r_1)$.
On the other hand, if the terminated connection is the other connection, its service rate is ${\mu _1}$.
Therefore, call termination associated with users in $U_{T_1^{(i)},T_1}$ happens with a rate of ${\cal N}_I^{(i)} [{\mu _1}( 1-r^{(i)}_I / r_1) + {\mu _1}]$. Notice that call recovery can always be successful in this case, since we utilize the recovery protection mechanism.

\subsubsection{Call terminated by users in $U_{T_1,T_2}$}
For call termination associated with users in $U_{T_1^{(i)},{T_2}}$ ($i \in \{0, 1,..., m_{II}\}$), we also should consider two cases.
When the ended call is the one in TFRC, ${\cal N}_{II}^{( i)} \to {\cal N}_{II}^{(i)} - 1$, ${n_2} \to {n_2} + 1$, with a rate of ${\cal N}_{II}^{(i)}  {\mu _1}  (1 - r_{II}^{(i)}/{r_1})$ (derived with the same approach used in Appendix A of \cite{Shan_TWC_submitted}).
When the ended call is the one without TFRC (i.e., the $T_2$-type connection), the released sunchannel number is $r_2$. However, to restore the spectrum supply to the connection in TFRC, $r_{II}^{(i)}$ subchannels are needed.
Note that $r_{II}^{(i)}$ may be larger than $r_2$ since $r_1 > r_2$.
It is thus possible that in this case the user cannot recover the recycled bandwidth (i.e., a recovering call dropping occurs) if $R(S_T) + r_{II}^{(i)} - {r_2} > C$.
Otherwise, call recovery can be successful, and we have ${n_1} \to {n_1} + 1$. The second case should occur with a rate of ${\cal N}_{II}^{(i)}{\mu _2}$.

%$T_2$ terminates with a rate of ${\cal N}_4^{(i)}{\mu _2}$, the released resources is $r_2$, while the resource required by the recovery of converted connection $T_1$ is $r_{II}^{(i)}$, $r_{II}^{(i)}$ may be larger than $r_2$ since $r_1 > r_2$. If $R(S_T) + r_{II}^{(i)} - {r_2} \le C$, ${\cal N}_4^{(i)} \to {\cal N}_4^{(i)} - 1$, ${n_1} \to {n_1} + 1$. Otherwise, the converted connection $T_1$ is dropped.

\subsubsection{Call terminated by users in $U_{T_2,T_1}$}

%Call termination associated with users in $U_{T_2,T_1}$ has similar  to
%
Specifically, when the ended call is associated with users in $U_{T_2^{(i)},T_1}$ ($i \in \{0, 1, ..., m_{III}\}$) and is the one in TFRC, ${\cal N}_{III}^{(i)} \to {\cal N}_{III}^{(i)} - 1$, ${n_1} \to {n_1} + 1$, with a rate of ${\cal N}_{III}^{(i)}  {\mu _2}  (1 - r_{III}^{(i)}/{r_2})$ (derived with the same approach used in Appendix A of \cite{Shan_TWC_submitted}).
On the other hand, when the ended call is the one without TFRC (i.e., the $T_1$-type connection), ${\cal N}_{III}^{(i)} \to {\cal N}_{III}^{(i)} - 1$, ${n_2} \to {n_2} + 1$, with a rate of $N_{III}^{(i)}{\mu _1}$.
However, as $r_1 > r_2$, call recovery can always be successful in this case.

%second opened connection $T_1$  terminates with a rate of $N_5^{(i)}{\mu _1}$, the resources released by $T_1$ are certainly sufficient for the recovery of converted connection $T_2$. Therefore, ${\cal N}_5^{(i)} \to {\cal N}_5^{(i)} - 1$, ${n_2} \to {n_2} + 1$.
%When a call belonging to a user of $U\left\{ {T_2^{(i)},{T_1}} \right\}$ terminates, there are also two cases:
%a. When the first opened and converted connection $T_2$ terminates with a rate of ${\cal N}_5^{(i)}{\mu _2}(1 - r_{III}^{(i)}/{r_2})$, ${\cal N}_5^{(i)} \to {\cal N}_5^{(i)} - 1$, ${n_1} \to {n_1} + 1$.
%b. when the second opened connection $T_1$  terminates with a rate of $N_5^{(i)}{\mu _1}$, the resources released by $T_1$ are certainly sufficient for the recovery of converted connection $T_2$. Therefore, ${\cal N}_5^{(i)} \to {\cal N}_5^{(i)} - 1$, ${n_2} \to {n_2} + 1$.

\subsubsection{Call terminated by users in $U_{T_2,T_2}$}
When the ended call is associated with users in $U_{T_2^{(i)},T_2}$ ($i \in \{0, 1, ..., m_{IV}\}$) and is the one in TFRC, ${\cal N}_{IV}^{(i)} \to {\cal N}_{IV}^{(i)} - 1$, ${n_2} \to {n_2} + 1$, with a rate of ${\cal N}_{IV}^{(i)}{\mu _2}(1 - r_{IV}^{(i)}/{r_2})$ (derived with the same approach in Appendix A of \cite{Shan_TWC_submitted}).
On the other hand, when the ended call is the one without TFRC, ${\cal N}_{IV}^{(i)} \to {\cal N}_{IV}^{(i)} - 1$, ${n_2} \to {n_2} + 1$, with a rate of ${\cal N}_{IV}^{(i)}{\mu _2}$.
Therefore, call termination associated with users in $U_{T_2^{(i)},T_2}$ happens with a rate of ${\cal N}_{IV}^{(i)}[{\mu _2}(1 - r_{IV}^{(i)}/{r_2}) + {\mu _2}]$.
However, call recovery can always be successful in this case.

%
%When a call belonging to a user of $U\left\{ {T_2^{(i)},{T_2}} \right\}$ terminates, there are also two cases:
%
%a. When the first opened and converted connection $T_2$ terminates with a rate of ${\cal N}_6^{(i)}{\mu _2}(1 - r_{IV}^{\left( i \right)}/{r_2})$, ${\cal N}_6^{(i)} \to {\cal N}_6^{(i)} - 1$, ${n_2} \to {n_2} + 1$.
%b. when the second opened connection $T_1$ terminates with a rate of ${\cal N}_6^{(i)}{\mu _2}$, ${\cal N}_6^{(i)} \to {\cal N}_6^{(i)} - 1$, ${n_2} \to {n_2} + 1$.
%
%Therefore, when a call belonging to a user of $U\left\{ {T_2^{(i)},{T_2}} \right\}$  terminates, ${\cal N}_6^{(i)}{\mu _2}$, ${\cal N}_6^{(i)} \to {\cal N}_6^{(i)} - 1$, ${n_2} \to {n_2} + 1$ with a rate of ${\cal N}_6^{(i)}{\mu _2}(1 - r_{IV}^{(i)}/{r_2}) + {\cal N}_6^{(i)}{\mu _2}$.

\subsection{A Periodic Time-Frequency Resource Conversion}

System state also changes with the periodic time-frequency resource conversion.
Specifically, at each feedback of context information, resource consumption of users with multiple connections can change.
Thus, when a new feedback instant arrives, the system state changes according to the TFRC strategy designed in Section III of \cite{Shan_TWC_submitted}: 1) the users of multiple connections but without TFRC apply their first round of spectrum reduction; and 2) the users of multiple connections with TFRC keep reducing spectrum supply until all subchannels of the connection in TFRC are widthdraw, leading to
%%
%For the users belonging to class 3 till class 6, all of them have two connections simultaneously, and the first opened connections are converted according to the means detailed above at every report of user terminal operating system. When the next report arrives:
%%
\begin{equation}%\label{}
    %\begin{array}{l}
        {\cal N}_j^{(0)} \to 0, {\cal N}_j^{(i)} \to {\cal N}_j^{(i-1)}, {\cal N}_j^{(m_j)} \to {\cal N}_j^{(m_j)} + {\cal N}_j^{(m_j - 1)} \notag%\\
   % \end{array}
\end{equation}
where $j \in \left\{I,II,III,IV\right\}$ and $i \in \left\{1, 2,..., {m_j } - 1\right\}$ correspond to one of the four types of users of multiple simultaneous connections and the state of the users' time-frequency resource conversion, respectively, all defined in \cite{Shan_TWC_submitted}.
%, $\Gamma  \in \left\{I, II, III, IV\right\}$.

%=======================================end of comments======================================

%%%%%%%%%%%%%%%%%%%%%%%%%%%%%%%%%%%%
% Acknowledgement
%%%%%%%%%%%%%%%%%%%%%%%%%%%%%%%%%%%%
%\section{Acknowledgement}
%\label{sec:Acknowledgement}

%The authors wish to thank the anonymous reviewers for their helpful reviews and suggestions which improved the quality and presentation of this paper.

%%%%%%%%%%%%%%%%%%%%%%%%%%%%%%%%%%%%
% Appendix
%%%%%%%%%%%%%%%%%%%%%%%%%%%%%%%%%%%%
%\vspace{-0.2cm}

%=======================================start comments==============================================

\section*{Appendix A: Handoff Rate Analysis}
\label{sec:Appendix_A}
In our model, call handoff can associate with different kinds of user. To derive handoff rate for each type of user, we analyze the user state transitions occurred in the system. Let us use the number of subchannels a user occupies to characterize the user  type.
Then the user state space can be given by
\begin{align}%\label{}
{\Upsilon _u} = %~~~~~~~~~~~~~~~~~~~~~~~~~~~~~~~~~~~~~~~~~~~~~~~~~~~~~~~~~~~~~~~~\notag \\
 \left\{ \begin{array}{l}
0,{r_1}, {r_2}, 2{r_1} - r_I^{(0)}, 2{r_1} - r_I^{(1)}, ... , 2{r_1} - r_I^{(m_I)},\\
{r_1} - r_{II}^{(0)} + {r_2}, {r_1}-r_{II}^{(1)} + {r_2}, ... ,{r_1} - r_{II}^{(m_{II})}+ {r_2},\\
{r_2} - r_{III}^{(0)} + {r_1}, {r_2} - r_{III}^{(1)} + {r_1}, ... , {r_2} - r_{III}^{(m_{III})} + {r_1} ,\\
2{r_2} - r_{IV}^{(0)}, 2{r_2} - r_{IV}^{(1)}, ..., 2{r_2} - r_{IV}^{(m_{IV})}
\end{array} \right\}
\end{align}
where without confusion we reuse the subchannel number to denote the user state. Specifically, here $0$, ${r_1}$, and ${r_2}$ denote the idle user, the user with a single $T_1$-type connection, and the user with a single $T_2$-type connection, respectively.
And $2{r_1} - r_I^{(i)}$, ${r_1} - r_{II}^{(i)} + {r_2}$, ${r_2} - r_{III}^{(i)} + {r_1}$, and $2{r_2} - r_{IV}^{(i)}$ respectively denote the user with two simultaneous $T_1$-type connections in which one is in state $S_1$ (foreground of the screen) and the other is in state $S_2$ (background of the screen),
the user with two simultaneous connections in which a $T_1$-type connection is in state $S_2$  and a $T_2$-type connection is in state $S_1$,
the user with two simultaneous connections in which a $T_2$-type connection is in state $S_2$  and a $T_1$-type connection is in state $S_1$,
and the user with two $T_2$-type connections in which one is in state $S_1$ and the other is in state $S_2$, with $i$ representing the progress of TFRC on the connection in state $S_2$.

Similar to system state, user state changes not only with call arrival and call termination but also with the periodic time-frequency resource conversion. Similar to \cite{Shan_TWC_submitted}, by dividing any state $S_u (\in {\Upsilon _u})$ into $M$ substates $S_{u}^{(m)}$, where $m= 1, 2, ..., M$, we can resort to multiple-stair Markov  model to approximate the mixed continuous-discrete Markov process of user state change \cite{Yu_MNA10,MarkovHTML}. Detailed analysis is omitted here, but the derived user state transitions are listed in Table II.
Based on the transition results we can obtain the set of  steady-state probabilities with respect to user state $\{ \pi (S_u^{(m)}) \}$, and further obtain $\left\{ \pi ({S_u}) \right\}$ because $\pi \left( {S_u} \right) = \sum\nolimits_{m = 1}^M {\pi (S_u^{(m)})}$.

As considered in \cite{Shan_TWC_submitted}, the total average user number in the cell is ${\bar K_A}$, %= \rho A$ where $\rho$ and $A$ are respectively the user density and the cell area,
and the residual time of user in the cell is exponentially distributed with mean $1/\eta$, then the handoff rate of the whole users in the cell is ${\bar K_A}\eta$.
Thus, the handoff rate of user with state $S_u$ can be found by taking the obtained  steady-state probability of the related user state into account, i.e., given by ${\bar K_A}\eta  \cdot \pi \left(S_u\right)$. It is noteworthy that, if there is only a single connection type in the system without applying TFRC, the handoff rate derived here can be proved consistent with  the result obtained in \cite{Kim_99}.

\begin{table*}
\centering
\renewcommand{\arraystretch}{0.75}
\caption{User state transitions.}
\vspace{-0.15cm}
\begin{tabular}{|lll|}
\hline
~~State transition & ~Relevant event & Transition rate \\ \hline\hline
$~~0\rightarrow r_{k},k=1,2$ &  ~The user without connection initiates a new $%
T_{k}$ connection  & $\lambda _{u}P_{k}$ \\ \hline
$~~r_{k}\rightarrow 2r_{k},k=1,2$ &  ~The user with a $T_{k}$  connection
initiates a new $T_{k}$ connection & $\lambda _{u}P_{k}$ \\ \hline
$~~r_{1}\rightarrow r_{1}+r_{2}$ &  ~The user with a $T_{1}$ connection
initiates a new $T_{2}$ connection & $\lambda _{u}P_{2}$ \\ \hline
$~~r_{2}\rightarrow r_{2}+r_{1}$ &  ~The user with a $T_{2}$ connection
initiates a new $T_{1}$ connection & $\lambda _{u}P_{1}$ \\ \hline
$~~r_{k}\rightarrow 0,k=1,2$ &  ~The single $T_{k}$ connection that the user owns
ends & $\mu _{k}$ \\ \hline
$%
\begin{array}{c}
2r_{1}-r_{I}^{(i)}\rightarrow r_{1} \\
i=0,1,...,m_{I}%
\end{array}%
$ & $%
\begin{tabular}{l}
If $i \neq m_I$, one of user's two $T_{1}$ connections (in which the one in \\
TFRC has been withdrawn $r^{(i)}_{I}$ subchannels) ends; otherwise, the \\
connection without TFRC ends%
\end{tabular}%
$ & $\mu _{1}(2-r_{I}^{(i)}/r_{1})$ \\ \hline
$%
\begin{array}{l}
r_{1}-r_{II}^{(i)}+r_{2}\rightarrow r_{1} \\
i=0,1,...,m_{II}%
\end{array}%
$ & $%
\begin{tabular}{l}
The $T_{2}$  connection that the user owns ends when the $T_{1}$ connection\\
in TFRC has been withdrawn $r_{II}^{(i)}$ subchannels%
\end{tabular}%
$ & $\mu _{2}$ \\ \hline
\begin{tabular}{l}
$r_{1}-r_{II}^{(i)}+r_{2}\rightarrow r_{2}$ \\
$i=0,1,...,m_{II}$%
\end{tabular}
&
\begin{tabular}{l}
If $i\neq m_{II}$, the $T_{1}$ connection ends when it has been
withdrawn $r_{II}^{(i)}$ \\
subchannels; otherwise, no state change occurs due to connection end%
\end{tabular}
& $\mu _{1}(1-r_{II}^{(i)}/r_{1})$ \\ \hline
$%
\begin{array}{l}
r_{2}-r_{III}^{(i)}+r_{1}\rightarrow r_{2} \\
i=0,1,...,m_{III}%
\end{array}%
$ & $%
\begin{tabular}{l}
The $T_{1}$ connection that the user owns ends when the $T_{2}$ connection \\
in TFRC has been withdrawn $r_{III}^{(i)}$ subchannels%
\end{tabular}%
$ & $\mu _{1}$ \\ \hline
$%
\begin{array}{l}
r_{2}-r_{III}^{(i)}+r_{1}\rightarrow r_{1} \\
i=0,1,...,m_{III}%
\end{array}%
$ &
\begin{tabular}{l}
If $i\neq m_{III}$, the $T_{2}$ connection ends when it has been
withdrawn $r_{III}^{(i)}$ \\
subchannels; otherwise, no state change occurs due to connection end%
\end{tabular}
& $\mu _{2}(1-r_{III}^{(i)}/r_{2})$ \\ \hline
$%
\begin{array}{l}
2r_{2}-r_{IV}^{(i)}\rightarrow r_{2} \\
i=0,1,...,m_{IV}%
\end{array}%
$ & $%
\begin{tabular}{l}
If $i \neq m_{IV}$, one of user's two $T_{2}$ connections (in which the one in \\
TFRC has been withdrawn $r^{(i)}_{IV}$ subchannels) ends; otherwise, the \\
connection without TFRC ends%
\end{tabular}%
$ & $\mu _{2}(2-r_{IV}^{(i)}/r_{2})$ \\ \hline
\begin{tabular}{l}
$(2r_{1}-r_{I}^{(i)})^{(M)}\rightarrow $ \\
$(2r_{1}-r_{I}^{(i+1)})^{(1)}$ \\
$i=0,1,...,m_{I}-1$%
\end{tabular}
&
\begin{tabular}{l}
Inter-state transition occurs at the beginning of a new TFRC period  \\
thus more subchannels are withdrawn from a $T_{1}$ connection  while\\
 the user focuses on the other $T_{1}$ connection%
\end{tabular}
& $M/\tau $ \\ \hline
\begin{tabular}{l}
$(r_{1}-r_{II}^{(i)}+r_{2})^{(M)}\rightarrow $ \\
$(r_{1}-r_{II}^{(i+1)}+r_{2})^{(1)}$ \\
$i=0,1,...,m_{II}-1$%
\end{tabular}
&
\begin{tabular}{l}
Inter-state transition occurs at the beginning of a new TFRC period \\
thus more subchannels are withdrawn from a $T_{1}$ connection while\\
 the user focuses on a $T_{2}$ connection%
\end{tabular}
& $M/\tau $ \\ \hline
\begin{tabular}{l}
$(r_{2}-r_{III}^{(i)}+r_{1})^{(M)}\rightarrow $ \\
$(r_{2}-r_{III}^{(i+1)}+r_{1})^{(1)}$ \\
$i=0,1,...,m_{III}-1$%
\end{tabular}
&
\begin{tabular}{l}
Inter-state transition occurs at the beginning of a new TFRC period \\
thus more subchannels are withdrawn from a $T_{2}$ connection while\\
 the user focuses on a $T_{1}$ connection%
\end{tabular}
& $M/\tau $ \\ \hline
\begin{tabular}{l}
$(2r_{2}-r_{IV}^{(i)})^{(M)}\rightarrow $ \\
$(2r_{2}-r_{IV}^{(i+1)})^{(1)}$ \\
$i=0,1,...,m_{IV}-1$%
\end{tabular}
&
\begin{tabular}{l}
Inter-state transition occurs at the beginning of a new TFRC period   \\
thus more subchannels are withdrawn from a $T_{2}$ connection while\\
 the user focuses on the other $T_{2}$ connection%
\end{tabular}
& $M/\tau $ \\ \hline
\begin{tabular}{l}
$(2r_{1}-r_{I}^{(m_{I})})^{(M)}\rightarrow $ \\
$(2r_{1}-r_{I}^{(m_{I})})^{(1)}$%
\end{tabular}
&
\begin{tabular}{l}
Intra-state transition occurs at the beginning of a new TFRC period \\
when the user has two $T_{1}$ connections in  which the one in TFRC \\
has been withdrawn all subchannels %
\end{tabular}
& $M/\tau $ \\ \hline
\begin{tabular}{l}
$(r_{1}-r_{II}^{(m_{II})}+r_{2})^{(M)}\rightarrow $ \\
$(r_{1}-r_{II}^{(m_{II})}+r_{2})^{(1)}$%
\end{tabular}
&
\begin{tabular}{l}
Intra-state transition occurs at the beginning of a new TFRC period \\
when the user has a $T_{1}$ connection and a $T_{2}$ connection in which\\
 the former is in TFRC and has been withdrawn all subchannels %at the end of a period of TFRC%
\end{tabular}
& $M/\tau $ \\ \hline
\begin{tabular}{l}
$(r_{2}-r_{III}^{(m_{III})}+r_{1})^{(M)}\rightarrow $ \\
$(r_{2}-r_{III}^{(m_{III})}+r_{1})^{(1)}$%
\end{tabular}
&
\begin{tabular}{l}
Intra-state transition occurs at the beginning of a new TFRC period \\
when the user has a $T_{1}$ connection and a $T_{2}$ connection in which\\
 the latter is in TFRC and has been withdrawn all subchannels %
\end{tabular}
& $M/\tau $ \\ \hline
\begin{tabular}{l}
$(2r_{2}-r_{IV}^{(m_{IV})})^{(M)}\rightarrow $ \\
$(2r_{2}-r_{IV}^{(m_{IV})})^{(1)}$%
\end{tabular}
&
\begin{tabular}{l}
Intra-state transition occurs at the beginning of a new TFRC period \\
when the user has two $T_{2}$ connections in which the one in TFRC\\
 has been withdrawn all subchannels
\end{tabular}
& $M/\tau $ \\ \hline
$~~0^{(M)}\rightarrow 0^{(1)}$ &
\begin{tabular}{l}
Intra-state transition occurs at the beginning of a new TFRC period \\
when the user has no connection
\end{tabular}
& $M/\tau $ \\ \hline
$~~r_{1}^{(M)}\rightarrow r_{1}^{(1)}$ &
\begin{tabular}{l}
Intra-state transition occurs at the beginning of a new TFRC period \\
when the user has only a $T_{1}$ connection%
\end{tabular}
& $M/\tau $ \\ \hline
$~~r_{2}^{(M)}\rightarrow r_{2}^{(1)}$ &
\begin{tabular}{l}
Intra-state transition occurs at the beginning of a new TFRC period \\
when the user has only a $T_{2}$ connection%
\end{tabular}
& $M/\tau $ \\ \hline
\begin{tabular}{l}
$S_{u}^{(m)}\rightarrow S_{u}^{(m+1)},\forall S_{u}\in \Upsilon _{u},$ \\
$m=1,2,...,M-1$%
\end{tabular}
&
\begin{tabular}{l}
Intra-state transition of any user state occurs in a temporal sequence \\
in a period of TFRC but before the end of a period of TFRC%
\end{tabular}
& $M/\tau $ \\ \hline
\end{tabular}
\end{table*}

\section*{Appendix B: Derivation of Call Holding Time Distribution of A Connection Applied TFRC}
\label{sec:Appendix_B}

Without loss of generality, take connection $j_1$ of Fig. 2 in \cite{Shan_TWC_submitted} as an example.
Let $l_{j_1}$ denote the call hold time of connection $j_1$ before applying TFRC, and $l'_{j_1}$ denote the call hold time of the connection successfully resumed full spectrum supply. Without loss of generality, we suppose that the spectrum recovery for $j_1$ occurs at the $(m+1)^
{th}$ context information feedback. Since the  data amount of the connection to be delivered remains the same with and without TFRC, we have
\begin{equation}%\label{}
{r_1} \cdot {R_b} \cdot (t_2 - t_1) + R_{rv}({j_1},{j_2},m) + {r_1}\cdot  {R_b}\cdot  {l'_{j_1}} = {r_1} \cdot {R_b} \cdot {l_{j_1}}
%\notag
\end{equation}
where the three items of the left side here are referred to as the delivered data amounts of connection $j_1$ in the following three disjointed durations, namely,
before the user gives his/her attention to the new connection $j_2$,
until the resource manager detects that the user refocuses on connection $j_1$,
and after the spectrum recovery for the connection.  %The second term $R_{rv}({j_1},{j_2},m)$ is given by (\ref{eq:rvdata}).
With  some manipulation, it is clear that
\begin{equation}%\label{}
{l'_{j_1}} = {l_{j_1}} - (t_2 - t_1) - R_{rv}({j_1},{j_2},m)/({r_1}{R_b}).%\notag
\end{equation}
Due to a positive value of call hold time, we have ${l_{j_1}} > (t_2 - t_1) + {R_{rv}}({j_1},{j_2},m)/({r_1}{R_b})$.
Thus, by applying the memoryless property of the exponential distribution, we have
\begin{equation}%\label{}
\begin{array}{l}
P\{\left. {l'_{j_1} > x} \right|l'_{j_1} > 0 \}
=P\left\{ {{l_{j_1}} - (t_2 - t_1) - \frac{R_{rv}(j_1,j_2,m)}{{r_1}{R_b}} > x} %\\
%~~~~~~~~~~~~~~~~~~~~~~~~~~~~~~
{\left| {{l_{j1}} > ({t_2} - {t_1}) + \frac{{{R_{{rv}}}({j_1},{j_2},m)}}{{r_1}{R_b}}} \right.} \right\}\\
~~~~~~~~~~~~~~~~~~~~~~~=P\{ l_{j_1} > x \}
\end{array}
\end{equation}
which implies that given  exponential distribution for the call hold time (${l_{j_1}}$) of a connection without applying TFRC,
as long as connection $j_1$ is successfully recovered, its call holding time  after spectrum recovery (${l'_{j_1}}$) follows the same distribution as the one that has not been applied TFRC.
%is a variable according to exponentially distributed and indicates the call holding time of $x_{j1}$ when TFRC is not carried out as denoted before. Thus ${l'_{I,j1}}$ is according to the same distribution as ${l_{j1}}$.

%=======================================end comments==============================================

\end{document}